\documentclass{aa}  

\usepackage{graphicx}
\usepackage{orcidlink}
\usepackage{txfonts}
\usepackage{mathtools}
\usepackage{tabularx}
\usepackage{booktabs}

\begin{document}

   \title{Improving Monte Carlo radiative transfer in the regime of high optical depths: The minimum scattering order}
    \titlerunning{Minimum scattering order}

   \author{A.~Krieger \orcidlink{0000-0002-3639-2435}
          \and
          S.~Wolf \orcidlink{0000-0001-7841-3452}
          }

   \institute{Institut für Theoretische Physik und Astrophysik, Christian-Albrechts-Universität zu Kiel, Leibnizstra{\ss}e 15, 24118 Kiel, Germany}

   \date{Accepted 6 November 2023}

 \abstract
 {Radiative transfer simulations are a powerful tool that enables the calculation of synthetic images of a wide range of astrophysical objects. These simulations are often based on the Monte Carlo method, as it provides the needed versatility that allows the consideration of the diverse and often complex conditions found in those objects. However, this method faces fundamental problems in the regime of high optical depths which may result in overly noisy images and severely underestimated flux values. In this study, we propose an advanced Monte Carlo radiative transfer method, namely, an enforced minimum scattering order that is aimed at providing a minimum quality of determined flux estimates. For that purpose, we extended our investigations of the scattering order problem and derived an analytic expression for the minimum number of interactions that depends on the albedo and optical depth of the system, which needs to be considered during a simulation to achieve a certain coverage of the scattering order distribution. The method is based on the utilization of this estimated minimum scattering order and enforces the consideration of a sufficient number of interactions during a Monte Carlo radiative transfer simulation. Moreover, we identified two notably distinct cases that shape the kind of complexity that arises in such simulations: the albedo-dominated case and the optical depth-dominated case. Based on that, we analyzed the implications related to the best usage of a stretching method as a means to alleviate the scattering order problem. We find that its most suitable application requires taking into account the value of the albedo as well as the optical depth. Furthermore, we argue that the derived minimum scattering order can be used to assess the performance of a stretching method with regard to the scattering orders its usage promotes. Finally, we stress the need for developing advanced pathfinding techniques to fully solve the problem of Monte Carlo radiative transfer simulations in the regime of high optical depths.
   }


   \maketitle
\section{Introduction}
Monte Carlo radiative transfer (MCRT) simulations are a widely used tool that enables the simulation and study of the physical conditions prevalent in various astrophysical systems. Among other applications, they allow us to calculate temperature distributions of these systems on the basis of numerical models, as well as to determine their appearance in terms of wavelength-dependent intensity maps. Due to its universality, this Monte Carlo (MC) based approach even allows for simulations of complex systems, such as filaments, protoplanetary or circumbinary disks, or planetary atmospheres. The basic principle behind MCRT codes such as RADMC-3D \citep{2012ascl.soft02015D}, POLARIS \citep{2016A&A...593A..87R}, or SKIRT\,9 \citep{2020A&C....3100381C} is the simulation of photon packages on randomly determined probabilistic paths. These photon packages are emitted from radiation sources and travel through the simulated model space, in which they interact potentially multiple times with the traversed medium before eventually leaving it. 

However, this approach faces a fundamental problem when simulating regions of high optical depths, as the required number of interactions and with it the computational demand rises. This ultimately puts a practical limit on the maximum optical depth that can be simulated. This well-known problem can result in noisy temperature distributions and flux values that are severely underestimated \citep[e.g.,][]{2016A&A...590A..55B,2017A&A...603A.114G,2018ApJ...861...80C,2020A&A...635A.148K}. Although it has already been shown that this issue may already arise for systems with optical depths between 10 and 30, where it has led to significant discrepancies in the outcome of different MCRT codes \citep{2017A&A...603A.114G}, the severity of it seems to increase strongly with the optical depth of the system, making the regime of high optical depths well-suited for the search for potential solutions.

Attempts have been made to alleviate this issue by the usage of advanced MCRT methods such as the partial diffusion approximation \citep{2009A&A...497..155M}, the modified random walk \citep{2009A&A...497..155M,2010A&A...520A..70R}, biasing techniques \citep{2016A&A...590A..55B}, precalculated sphere spectra \citep{2020A&A...635A.148K}, and the extended peel-off method \citep{2021A&A...645A.143K}. Each of these methods has their respective use cases, nonetheless, the core problem remains unsolved. For this reason, \citet{2018ApJ...861...80C} carried out a study, focussing on the phenomenon of underestimated flux values in the regime of high optical depths. They analyzed the limit of optical depth values that can be simulated on the basis of the numerical setup of an infinite plane-parallel slab that is illuminated from one side, for which they determined the transmitted intensity using different MCRT methods. According to this study, the most reliable estimate for the transmitted intensity was achieved, when utilizing a combination of different MCRT methods. Their best method, in particular, used a split method, which reduces the weight of a photon package at each point of interaction, but in return enables the enforcement of the interaction to be a scattering event rather than an absorption event. Additionally, a stretching method was applied, which biases the distance between two consecutive points of interaction toward larger optical depths, i.e., it stretches the path of the photon package. 

\citet{2021A&A...645A.143K} then performed an in-depth study on the core problems of MCRT simulations in the regime of high optical depths and identified the scattering order problem, which is accompanied by a pathfinding problem. The scattering order problem can be summarized as follows: The total simulated flux is composed of contributions from simulated photon packages of different scattering orders; the latter refers to the number of scattering events a photon package has undergone prior to its simulated detection. A reliable flux estimate, thus, relies on a proper representation of the scattering order distribution (SOD), namely, the distribution of contributions to the total flux with regard to the different scattering orders. At higher optical depths, however, this is becoming increasingly difficult, as the range of relevant scattering orders may broaden and shift toward extremely large values \citep{2021A&A...645A.143K}, resulting in the need for the simulation of a large quantity of photon packages of potentially extremely high scattering orders. This problem was shown to be alleviated with the additional utilization of the peel-off method \citep{1984ApJ...278..186Y} or the extended peel-off method \citep{2021A&A...645A.143K}; nonetheless, (transverse) optical depth values of $\tau_{\max}\gtrsim 150$ still resulted in a notable underestimation of transmitted flux values when restricting the simulation time to a feasible extent. This breakdown at high optical depths could, yet again, be attributed to an inadequate representation of the SOD, stressing the need for more efficient MCRT methods. Thus, apart from a potential improvement of the method that is used to determine photon package paths, the question of the smallest number of required simulated scattering events was raised. In MCRT simulations, the total flux strictly increases with the number of considered scattering events per photon package, assuming the simultaneous usage of a peel-off based method; flux values may therefore be severely underestimated, if the simulated photon packages do not undergo a certain minimum number of scattering events. 

The concept of such a minimum scattering order is the subject of the present study. In Sect. \ref{sec:derivation}, we derive an approximate SOD for the testbed of a one-dimensional (1D) slab and we explore two cases that shape the kind of difficulty that arises in the regime of high optical depth: the albedo-dominated case and the optical depth-dominated case. Based on that, we present a simple formula in Sect. \ref{sec:minimum_scattering_order} that estimates the required minimum scattering order and we propose a simple method that allows for its utilization. In Sections \ref{sec:application_slab} and \ref{sec:application_disk} the proposed method is applied to assess its benefits based on MCRT simulations of two different systems, a three-dimensional (3D) infinite plane-parallel slab and a protoplanet embedded in an edge-on circumstellar disk, respectively. In Sect. \ref{sec:discussions}, the limits of the method are discussed, along with the implications of our findings regarding the effectiveness of stretching methods and a potential way for their assessment. Finally, our results are summarized in Sect. \ref{sec:summary}.

\section{Derivation}
\label{sec:derivation}

In order to explore the possibility for improving the capabilities of modern MCRT simulations at high optical depths, we studied the setup of a homogeneous 1D slab, which is illuminated from one side by a monochromatic radiation source and we analyzed the SOD of photons, that are represented by photon packages in a MCRT simulation, transmitted through the slab. This study builds upon the results of our previous work \citep{2021A&A...645A.143K}, which we summarize in Sect. \ref{sec:setup}. In Sect. \ref{sec:SOD} we derived an approximate solution for the underlying SOD of the studied setup, which we use to infer various of its properties. In Sect. \ref{sec:complexity}, we discuss these findings with regard to the implications regarding the complexity of MCRT simulations, before proposing a simple to implement method that aims at solving the problems at high optical depths in Sect. \ref{sec:minimum_scattering_order}. 

\subsection{Setup}
\label{sec:setup}

We adopted the setup of \citet{2021A&A...645A.143K}, which is composed of a homogeneous slab with a total (extinction) optical depth of $\tau_{\rm max}$ that is embedded in a vacuum. For any given distribution $\nu_0(x)$ of photons (or photon packages) inside the slab, where $x\in \mathbb{R}$ describes the position in terms of optical depth units with $x=0$ the center of the slab, the radiative transfer process can be described by an iterative process. During its $n$-th step, photons are launched at locations $x$, according to their current distribution $\nu_n(x)$, and traverse the slab until either leaving it through one of its borders or interacting at a different location within the slab. This results in the distribution $\nu_{n+1}(x)$ of the next step. The randomly selected traversed distances between two consecutive points of interaction follow an exponential distribution, resulting in a spreading of the photon distribution during each step, while the directions are chosen randomly according to an isotropic distribution. The event of interaction itself is simulated by rescaling the whole photon distribution by a factor of $A$, which corresponds to the single scattering albedo of the slab. The process of spreading can be described in terms of eigenstates \citep{2021A&A...645A.143K}, which themselves are photon distributions that remain unchanged during this process, apart from some eigenstate-specific constant factor $\lambda$, namely, the eigenvalue. To distinguish between symmetric and anti-symmetric eigenstates, we marked variables corresponding to the latter with an asterisk. For more details, we refer to the appendix in \citet{2021A&A...645A.143K}.

   \begin{figure}
   \centering
   \includegraphics[width=\hsize]{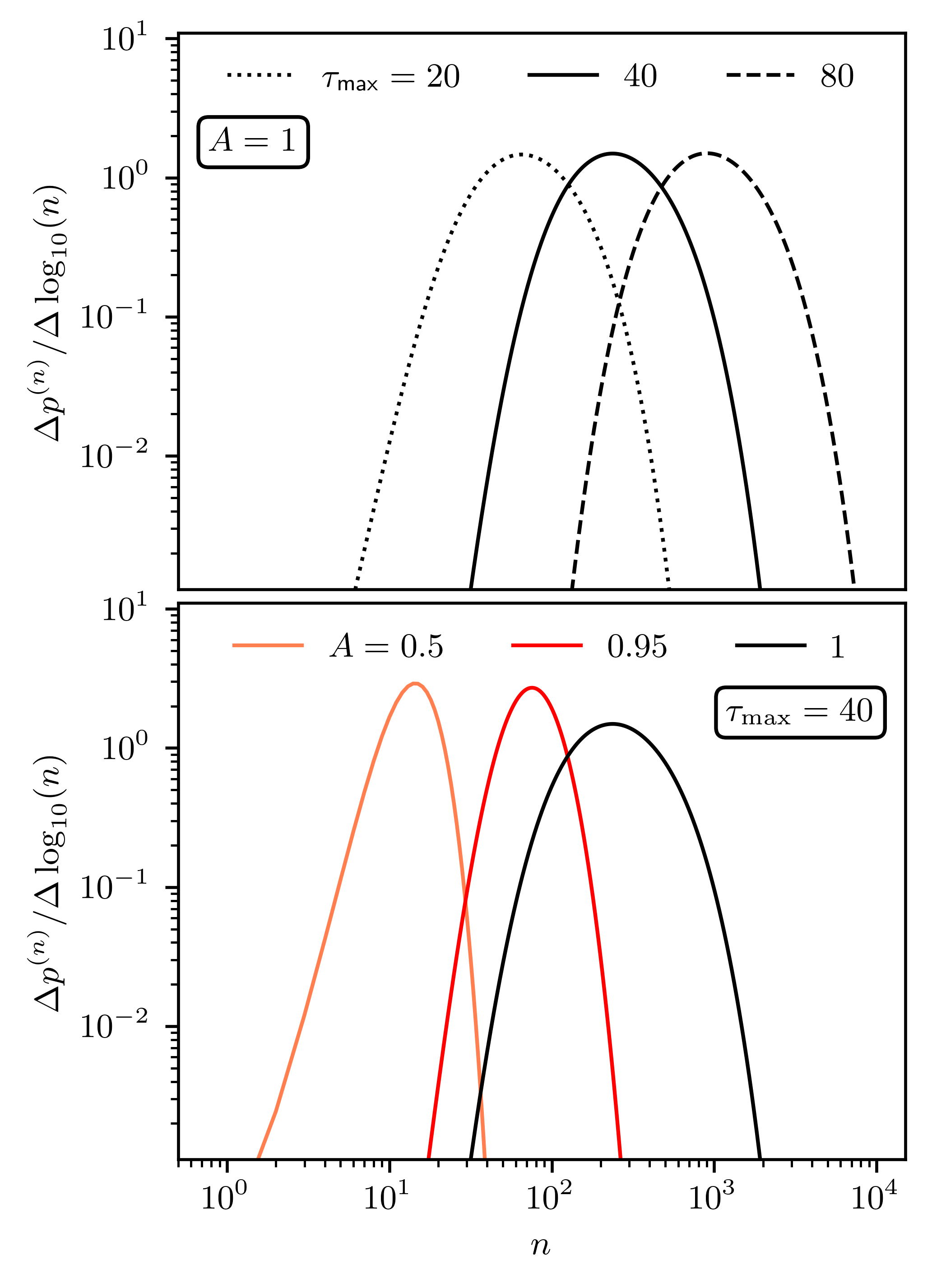}
      \caption{The SOD $\Delta p^{(n)}/\Delta \log_{10}{(n)}$ as function of the scattering order $n$. \emph{Top:} Impact of $\tau_{\rm max}$ using $A\!=\!1$.\, \emph{Bottom:} Impact of $A$ using $\tau_{\rm max}\!=\!40$.}
         \label{fig:scattering_order_distribution}
   \end{figure}

The total intensity of photons that escape the slab on one side can then be described in terms of a weighted sum of all eigenstates of the slab with eigenvalues $\lambda_k$ and $\lambda_k^*$ with $k\in\mathbb{N}$ and the albedo $A$. The contribution $I^{(n)}$ of the $n$-th scattering order to the total intensity is given by \citep{2021A&A...645A.143K}
\begin{equation}
    I^{(n)} = A^n \sum_{k=1}^{\infty} \left( -1 \right)^{k+1} \left( \lambda_k^{n+1} \sqrt{1-\lambda_k}\, a_k + {\lambda_k^*}^{n+1}\sqrt{1-\lambda_k^*}\, b_k \right).
    \label{eq:I_n_general}
\end{equation}
The coefficients $a_k$ and $b_k$ of this expansion can be derived from an initial photon distribution, hereby given as:
\begin{equation}
    \nu_0(x) = 2 \delta \left( x+b\right). \label{eq:initial_dist}
\end{equation}
This particular distribution was chosen since it represents a scenario in which the slab is illuminated from one side by a monochromatic radiation source. To this end, the source is placed on one side of the slab at the optical depth coordinate $x=-b$ with $b=\tau_{\rm max}/2$. Furthermore, it is normalized due to a factor of 2 in Eq. \eqref{eq:initial_dist}, which ensures that the totality of all simulated photons that traverse the slab after the first event of emission carry a weight of one, as the underlying phase function follows an isotropic distribution. The coefficients $a_k$ and $b_k$ in Eq. \eqref{eq:I_n_general} are then given by \citep{2021A&A...645A.143K}
\begin{equation*}
    a_k = \left\langle \cos{(\sigma_k x)},\,\nu_0(x) \right\rangle = \frac{2 \cos{(\sigma_k b)}}{\lambda_k + b} = (-1)^{k+1}\,\frac{2 \sqrt{1-\lambda_k}}{\lambda_k + b} 
\end{equation*}
and
\begin{equation*}
    b_k = \left\langle \sin{(\sigma_k^* x)},\,\nu_0(x) \right\rangle = \frac{-2 \sin{(\sigma_k^* b)}}{\lambda_k^* + b} = (-1)^{k}\,\frac{2 \sqrt{1-\lambda_k^*}}{\lambda_k^* + b}. 
\end{equation*}
In combination with Eq. \eqref{eq:I_n_general} this results in the following expression for $I^{(n)}$:
\begin{equation}
    I^{(n)} = 2 A^n \sum_{k=1}^{\infty}  \frac{\lambda_k^{n+1}\left(1-\lambda_k\right)}{\lambda_k + b} -\frac{{\lambda_k^*}^{n+1}\left(1-\lambda_k^*\right)}{ \lambda_k^* + b}.
    \label{eq:I_n_sum}
\end{equation}
This expression can then be used to calculate the probability $\Delta p^{(n)} = I^{(n)}/\sum_{n=0}^\infty I^{(n)}$ for a transmitted photon package to have scattered $n$ times before leaving the slab, and with it the SOD, namely, the distribution of scattering orders $n$ of transmitted photon packages. Figure \ref{fig:scattering_order_distribution} exemplarily shows the SODs $\Delta p^{(n)}/\Delta \log_{10}{(n)}$ for three different values of the optical depth $\tau_{\rm max}$ (upper plot) and for three different value of the albedo $A$ (lower plot), which were calculated as described in \citet{2021A&A...645A.143K}. These plots clearly show a shift of the SOD toward higher values of the scattering order $n$ for increasing values of optical depth and albedo, which will be addressed in greater detail later on in this study. In the following sections, we derive an approximate solution for the SOD and discuss its implications for the complexity of MCRT simulations in the regime of high optical depths and the related scattering order problem. The results of these derivations will form the basis for a novel proposed method that aims at solving or at least alleviating this problem and should thus improve MCRT simulations in the regime of high optical depths.

\subsection{The scattering order distribution}
\label{sec:SOD}
Every eigenvalue $\lambda$ is related to its corresponding (positive) $\sigma$ value, in accordance with the equation $\lambda = 1/(1+\sigma^2)$, which can be obtained by solving the equations \citep{2021A&A...645A.143K}
\begin{eqnarray}
    b \sigma_k &=& \pi \left( k-1\right) + b \Delta \sigma_k \label{eq:b_sigma_k} \quad \text{and}\\
    b \sigma_k^* &=& \pi \left( k-0.5\right) + b \Delta \sigma_k^* \label{eq:b_sigma_k_star}
\end{eqnarray}
for $k\in \mathbb{N}$, where
\begin{equation}
    b \Delta \sigma_k = \tan^{-1}(\sigma_k^{-1}) \quad \text{and} \quad b \Delta \sigma_k^* = \tan^{-1}({\sigma_k^*}^{-1}). \label{eq:b_delta_sigma_k_star}
\end{equation}
It can be shown, that the inequalities $0 < b \Delta \sigma_{k+1} < b \Delta \sigma_k < \pi/2$ and $0 < b \Delta \sigma_{k+1}^* < b \Delta \sigma_k^* < \pi/2$ hold, meaning that both $b \Delta \sigma_{k}$ and $b \Delta \sigma_k^*$ decrease with increasing value of $k$. We note that previous equations imply that quantities with asterisks are related to the same quantities without asterisks according to $\lambda^*_k=\lambda_{k+0.5}$ and $\sigma^*_k=\sigma_{k+0.5}$. However, in order to use these relations, we have to extend $\sigma_k$ and $\sigma_k^*$ as well as $\lambda_k$ and $\lambda_k^*$ to real numbers of $k$. Using Eqs. \eqref{eq:b_sigma_k}--\eqref{eq:b_delta_sigma_k_star} and performing derivatives with respect to $k$, the extension is defined by the derivative
\begin{equation*}
    \frac{{\rm d} \sigma}{{\rm d}k} = \frac{\pi}{\left(1+\sigma^2\right)^{-1} + b} 
\end{equation*}
for $\sigma\in \left\{\sigma_k, \sigma_k^*\right\}$ as well as
\begin{equation}
    \frac{{\rm d} \lambda}{{\rm d}k} = \frac{-2\pi\sqrt{1-\lambda}\, \lambda^{3/2}}{\lambda + b}\label{eq:dlambda_dk}
\end{equation}
for $\lambda\in \left\{\lambda_k, \lambda_k^*\right\}$. For a fixed value of $\lambda$, the derivative in Eq. \eqref{eq:dlambda_dk} is on the order of $1/b$ in the regime of $b\gg 1>\lambda$, while the second derivative $\frac{{\rm d}^2 \lambda}{{\rm d}k^2}$ is on the order of $1/b^2$. In this regime, the difference $\Delta \lambda_k=\lambda_{k+1} -\lambda_{k}$ between two consecutive $\lambda$ values can thus be approximated by 
\begin{equation*}
    \Delta \lambda_k \approx \frac{{\rm d} \lambda}{{\rm d}k}\, \Delta k, 
\end{equation*}
with $\Delta k = 1$. Using this relation, the series in Eq. \eqref{eq:I_n_sum} becomes
\begin{equation}
    I^{(n)} = - \sum_{k=1}^{\infty}  U^{(n)}(\lambda_k)\, \Delta \lambda_k - U^{(n)}(\lambda_k^*)\, \Delta {\lambda_k^*}, \label{eq:sum_to_integral}
\end{equation}
with 
\begin{equation*}
    U^{(n)}(\lambda) \equiv \frac{A^n}{\pi} \lambda^{n-0.5}\sqrt{1-\lambda},
\end{equation*}
such that $I^{(n)}$ converges in the limit of $b\rightarrow \infty$ to the difference of two Riemann integrals, one performed over $\lambda_k$ and another over $\lambda_k^*$. In particular, the upper boundary of the integration intervals of both integrals would differ by $\Delta \lambda = \lambda_1-\lambda_1^*$, while the lower boundary for both these integrals would be zero and thus equivalent. By approximating the right-hand side of Eq. \eqref{eq:sum_to_integral} with this difference, we arrive at
\begin{equation}
    I^{(n)} \approx U^{(n)}(\lambda_1)\left( \lambda_1-\lambda_1^*\right) \label{eq:In_approx_using_lambda0}.
\end{equation}
Since the scattering order, $n$, appears as an exponent in the function $U^{(n)}$, it can be shown that the approximation in Eq. \eqref{eq:In_approx_using_lambda0} is underestimating the actual intensity at very high scattering orders. Moreover, in the case of very low scattering orders, we find that the approximation overestimates the flux, suggesting that the approximate SOD is generally shifted toward lower scattering orders compared to its exact solution. This suggests that any derived expression for a minimum scattering order can be expected to be rather underestimated. This is discussed in the following sections.

Finally, the determination of $I^{(n)}$ requires the calculation of $\lambda_1$ and $\lambda_1^*$, which can be deduced from their corresponding $\sigma$ values, $\sigma_1$ and $\sigma_1^*$, respectively. 
In the regime of high optical depths, the conditions $0<\sigma_1 \ll 1$ and $0<\sigma_1^* \ll 1$ are satisfied. Thus, we perform a series expansion of Eqs. \eqref{eq:b_sigma_k} and \eqref{eq:b_sigma_k_star} about $\sigma=0$ to deduce both values as functions of the optical depth $b$. This results in equations of the form:
\begin{equation*}
    b\sigma = f\pi - \sigma + \frac13 \sigma^3 + \mathcal{O}(\sigma^5), 
\end{equation*}
where $f=0.5$ for $\sigma = \sigma_1$ and $f=1$ for $\sigma = \sigma_1^*$. In the considered optical depth regime, this cubic equation has three real roots, $\sigma_A$, $\sigma_B$, and $\sigma_C$, which satisfy $\sigma_A < 0 <\sigma_B < \sqrt{b+1}< \sigma_C$, with $\sigma_B$ being the searched-for solution. It can be shown that its solution is given by
\begin{equation*}
    \sigma_B = \frac{f\pi}{b} - \frac{f\pi}{b^2} + \frac{f\pi}{b^3} + \frac{f\pi \left( (f\pi)^2/3 -1 \right)}{b^4} + \mathcal{O}\left( \frac{1}{b^5}\right).
\end{equation*}
Making use of the relation between $\sigma$ and $\lambda$ values, we find
\begin{eqnarray*}
    \lambda_1 &=& 1 - \frac{\pi^2}{4b^2} + \frac{\pi^2}{2b^3} + \mathcal{O}\left( \frac{1}{b^4} \right) \quad \text{and}\\
    \lambda_1^* &=& 1 - \frac{\pi^2}{b^2} + \frac{2\pi^2}{b^3} + \mathcal{O}\left( \frac{1}{b^4} \right).
\end{eqnarray*}
Finally, using these results in combination with Eq. \eqref{eq:In_approx_using_lambda0} and neglecting higher order contributions, we arrive at an approximate solution for the SOD for photons, which traversed the slab in the regime of high optical depths, that is given by
\begin{equation}
    I^{(n)} =  \frac{3 \pi^2}{8 b^3} \left(A \lambda_1 \right)^n \label{eq:In_using_exp_approximation}.
\end{equation}

\subsection{Complexity of radiative transfer simulations}
\label{sec:complexity}

The approximate solution for $I^{(n)}$ shown in Eq. \eqref{eq:In_using_exp_approximation} for the regime of high optical depths can then be used to calculate the position of the maximum of the SOD, $n_{\rm max}$, which satisfies 
\begin{equation*}
    \left.\frac{{\rm d}}{{\rm d}n} \left( nI^{(n)}\right)\right\rvert_{n=n_{\rm max}}=0
\end{equation*}
and is thus given by 
\begin{equation}
    n_{\rm max}=\left.\frac{-I^{(n)}}{{\rm d}I^{(n)}/{\rm d}n}\right\rvert_{n=n_{\rm max}} = \frac{-1}{\ln{A}+\ln{\lambda_1}}. \label{eq:n_max}
\end{equation}
Equations \eqref{eq:In_using_exp_approximation} and \eqref{eq:n_max} suggest that while in the regime of high optical depths the transmitted flux decreases strongly with the optical depth, the complexity of the radiative transfer process can still be limited by the albedo rather than the optical depth. Here, we refer to the complexity in terms of the scattering order ranges that need to be simulated to properly represent the radiative transfer process, which is particularly important for deriving reliable flux estimates. This implies that the most suitable MCRT method for solving a specific problem should always consider the complexity that is introduced due to the albedo and the optical depth. 

In particular, if the complexity is dominated by the optical depth, namely, if $|\ln{A}|\ll |\ln{\lambda_1}|$, the position of the SOD becomes practically independent of the albedo and is determined by the optical depth according to 
\begin{equation}
    n_{\rm max}\approx -1/\ln{\lambda_1}\rightarrow 4b^2/\pi^2 + \mathcal{O}(b) \hfill\text{(optical depth-dominated)}. \label{eq:nmax_optical_depth_dominated}
\end{equation}

Comparing this result with the SODs in the upper plot of Fig. \ref{fig:scattering_order_distribution}, we find that the actual position of the maximum exhibits an approximately quadratic dependency on the optical depth, as predicted by Eq. \eqref{eq:nmax_optical_depth_dominated}. However, we also find, that the derived approximation underestimates the scattering order of the maxima, as was expected as a result of a shift of the approximate solution toward lower scattering orders (see Sect. \ref{sec:SOD}). Furthermore, we note that the special case of $A=1$ is particularly interesting, in the sense that it always results in an optical depth-dominated scenario, independent of the actual value of $\tau_{\rm max}$, and thus the complexity of the problem increases drastically with the optical depth. 

On the contrary, if the complexity of the radiative transfer process is albedo-dominated, namely, if $|\ln{\lambda_1}|\ll |\ln{A}|$, the position of the maximum is mostly determined by the albedo and given by 
\begin{equation}
    n_{\rm max}\approx -1/\ln{A}\hfill\text{(albedo-dominated)}. \label{eq:nmax_albedo_dominated}
\end{equation}
Similarly to the optical depth-dominated case, we find that this approximate position of the maximum also underestimates the scattering order of the actual position of the maximum for the SODs shown in the lower plot of Fig. \ref{fig:scattering_order_distribution}. This suggests that the derived expressions for $n_{\rm max}$ in Eq. \eqref{eq:nmax_optical_depth_dominated} and \eqref{eq:nmax_albedo_dominated} can be used as lower boundaries for the actual position of the maximum of an SOD in both the optical depth-dominated and the albedo-dominated cases, respectively. 

Another property that strongly impacts the complexity is the mean number of scattering orders $n_{\rm mean}$, which is given by
\begin{equation*}
    n_{\rm mean} = \frac{\sum_{n=0}^\infty n I^{(n)}}{I^{\rm tot}} = \frac{A\lambda_1}{1- A\lambda_1},
\end{equation*}
where $I^{\rm tot}$ is the total transmitted intensity\footnote{We note that, according to Eq. \eqref{eq:In_using_exp_approximation}, $I^{\rm tot}$ decreases $\propto 1/b$ in the optical depth-dominated case, for instance if $A=1$, while it is shown to be falling off $\propto 1/b^3$ in the albedo-dominated case. Thus, for any fixed value of $A\lesssim1$, the slope of the total transmitted intensity as function of the optical depth decreases from -1 to -3 in the regime of high optical depths.} with:
\begin{equation*}
    I^{\rm tot} = \sum_{n=0}^\infty I^{(n)}.
\end{equation*}
We find, that $n_{\rm mean}$ is closely related to the position of the peak of the SOD $n_{\rm max}$, which satisfies $A\lambda_1 \leq n_{\rm mean}/n_{\rm max} \leq 1$. In the optical depth-dominated case, the position of the peak of the SOD therefore approximately traces the mean number of scattering orders. 

Furthermore, the width of the SOD also strongly impacts the complexity of the radiative transfer problem. Using the approximation for $I^{(n)}$ in Eq. \eqref{eq:In_using_exp_approximation}, we find that the apparent width of the SOD in the double logarithmic plot of the SOD (see Fig. \ref{fig:scattering_order_distribution}) indeed converges to a constant value in the regime of high optical depths, as it has been conjectured in \citet{2021A&A...645A.143K}. Considering the results of this study regarding the width of the SOD and position of its maximum as a function of the albedo and optical depth, we therefore find more evidence in support of the importance of the scattering order problem \citep{2021A&A...645A.143K}, suggesting that it indeed poses one of the fundamental problems of MCRT simulations in the regime of high optical depths.

\section{Method: Minimum scattering order}
\label{sec:minimum_scattering_order}

In the regime of high optical depths, it is crucial to simulate photon packages of high scattering orders to achieve a proper representation of the actual SOD and, thus, a reliable flux estimate. Photon packages that leave the simulated model space with a very small scattering order cannot contribute toward a sufficient coverage of the SOD and may thus lead to a misrepresentation of the distribution and the underestimation of flux \citep{2021A&A...645A.143K}. 

One possibility to counteract this is to enforce a minimum scattering order, $m$, which we recommend coupling with the usage of a peel-off method, meaning, photon packages may only leave the model space after their number of scattering events reached or even exceeded this threshold. Using Eq. \eqref{eq:In_using_exp_approximation}, we arrive at the following relation between the coverage $p\in\left[0,1 \right)$ of the total transmitted intensity and the approximate scattering order $m$, that needs to be simulated in order to reach it:
\begin{equation*}
    p = \frac{\sum_{n=0}^{m}I^{(n)}}{I^{\rm tot}} = 1 - \left(A\lambda_1 \right)^{m+1}
\end{equation*}
Hence, in the regime of high optical depths the minimum scattering order, as a function of the albedo $A>0$, optical depth, and coverage $p$, is given by
\begin{equation}
    m = \frac{\ln{\left( 1-p \right)}}{\ln{\left( A\lambda_1 \right)}} - 1 \stackrel{\text{Eq. \eqref{eq:n_max}}}{=} -\ln{\left( 1-p \right)}\,n_{\rm max}  - 1.\label{eq:m_general}
\end{equation}
When assuming a coverage of $p=99.9\%$, this results in the following expressions for the minimum scattering order for both the albedo-dominated and the optical depth-dominated cases:
\begin{equation}
    m \approx 
    \begin{dcases} 
        \frac{28 b^2}{\pi^2} - 1   \hfill\hspace{.3cm}\text{(optical depth-dominated; $A=1$) }\\
        \frac{-7}{\ln{A}} - 1  \hspace{.6cm}\text{(albedo-dominated; $A<1$ and $b\rightarrow \infty$).}
    \end{dcases}
    \label{eq:m_999}
\end{equation}
In order to apply our proposed method of enforced minimum scattering order during a MCRT simulation, simulated photon packages are kept inside the model space for a minimum of $m$ scattering events, where $m$ is chosen at least as high as stated in Eq. \eqref{eq:m_999}. During a standard MCRT simulation, an arbitrarily large optical depth $\tau$ is chosen after every single event of interaction randomly according to its corresponding probability distribution to decide the next location of interaction. However, when using our proposed method, instead the interval is limited to values of $\tau<\tau_{\rm los}$, where $\tau_{\rm los}$ is the total optical depth in the direction of the original photon package. This enforces an interaction within the model space and to compensate for that, the weight of the resulting photon package needs to be adjusted by a factor of $w_{\rm enforce} = 1-e^{-\tau_{\rm los}}$. This factor corresponds to the probability with which the original photon package interacts at an optical depth of $\tau<\tau_{\rm los}$. 

We note that this method furthermore requires the simulation of a secondary, non-interacting photon package to compensate for the loss of weight carried by the original photon package. Its direction corresponds to that of the original photon package, while its weight is equal to that of the original package, reduced by a factor of $w_{\rm sec}=e^{-\tau_{\rm los}}$. The simulation of such a secondary photon package, however, is no longer required if a peel-off method is used simultaneously, since in this case its weight must not add to the flux estimate, rendering its simulation unnecessary. Additionally, if the peel-off method is used, every simulated event of interaction contributes to the SOD, leading to an improvement of the overall coverage of scattering orders.

\subsection{Application: Slab experiment}
\label{sec:application_slab}
In this section, we apply our proposed method in a simulation of an infinite plane-parallel 3D homogeneous slab. This setup is often used as a testbed, particularly for the study of MCRT simulations in the regime of high optical depths \citep[for details regarding the setup, see][]{2018ApJ...861...80C,2021A&A...645A.143K}. The slab is illuminated from one side by an isotropic monochromatic radiation source, where the slab has a total transverse optical depth of $\tau_{\rm max}=30$. In this setup, we simulate the transmitted intensity as a function of $\mu = \cos{\theta}$, which is the cosine of the penetration angle $\theta$. For this test, the single scattering albedo is set to $A=0.7$ and the scattering of photon packages is simulated according to an isotropic phase function. To estimate the overall performance of the MCRT simulations, we additionally calculate the expected transmitted intensity based on a non-probabilistic solution. To this end, we adopted the procedure used by \citet{2018ApJ...861...80C} and \citet{2021A&A...645A.143K} that is based on work done by \citet{1983ApJ...275..292R} and \citet{1995MNRAS.277.1279D}.

   \begin{figure}
   \centering
   \includegraphics[width=\hsize]{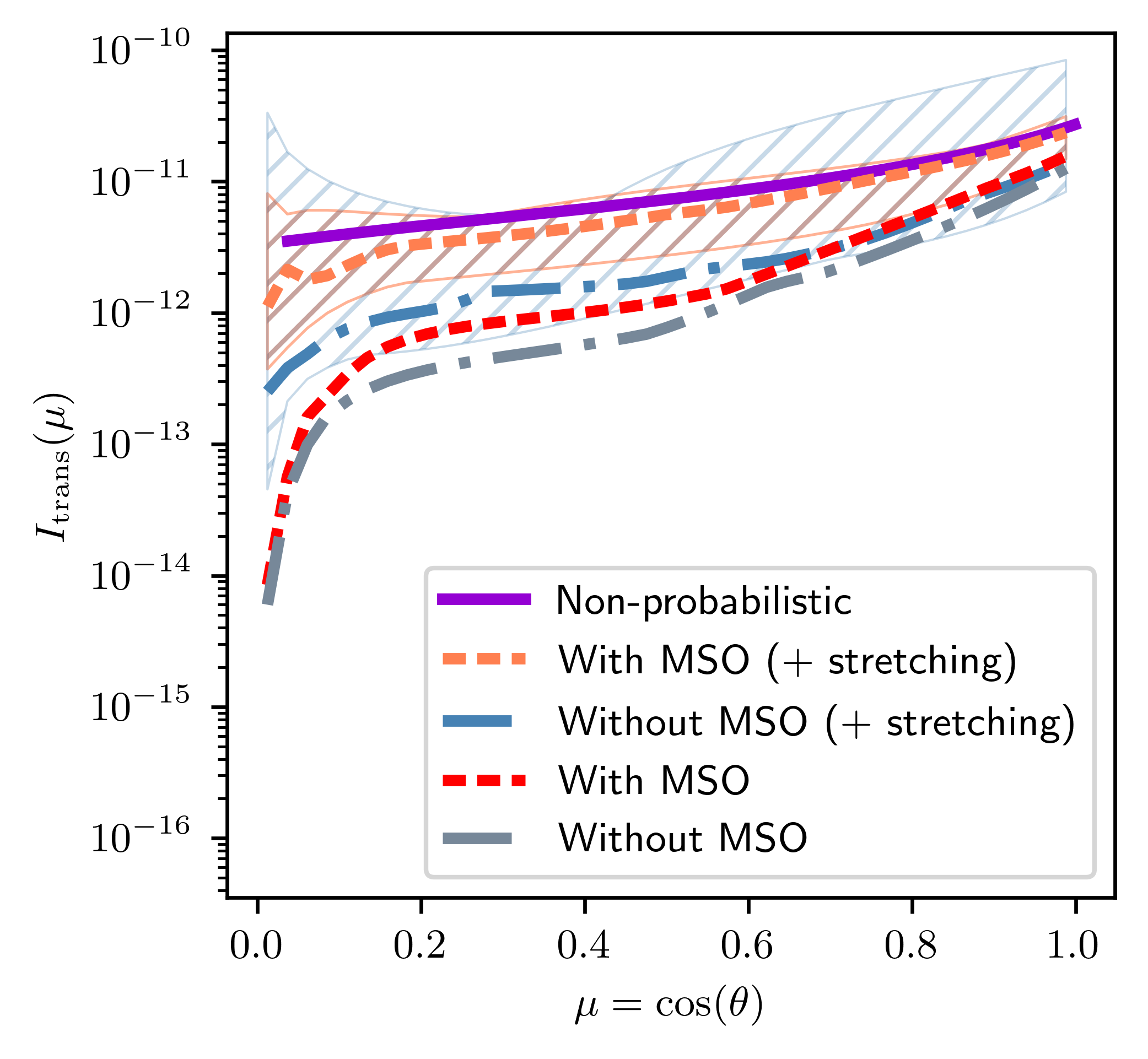}
      \caption{Results for the infinite plane-parallel slab experiment for an optical depth of $\tau_{\rm max}=30$ and an albedo of $A=0.7$. Curves represent transmission intensities obtained by either a non-probabilistic method (purple) or MC based methods (dashed curves). The latter are additionally divided into four types, depending on their usage of advanced MCRT methods, such as the stretching method (stretch) or the enforcement of a minimum scattering order (MSO). The underlaid colored areas mark the regions of all simulated intensity curves, and dashed curves of the same color represent the corresponding medians.}
         \label{fig:comparison_Itrans_tau30}
   \end{figure}

Figure \ref{fig:comparison_Itrans_tau30} shows results for the determined transmitted intensity $I_{\rm trans}(\mu)$ for the non-probabilistic solution (purple solid line) as well as for four different types of MC based radiative transfer simulations. The latter differ with regard to their utilization of our proposed minimum scatter order method (MSO) as well as regarding the usage of the composite-biasing technique \citep{2018ApJ...861...80C} for stretching photon package paths. For each of these types of simulations, we performed $10^4$ simulations with $10^3$ photon packages per simulation for our further analysis. These data were subsequently divided into ten batches of $10^3$ simulations each, across which the average transmitted intensity was calculated. This effectively yields intensity curves corresponding to $N=10^6$ simulated photon packages each. Then, the four displayed dashed curves represent the respective median curves of all ten thereby obtained intensity curves, while the hatched colored areas mark regions in the plot, where all individual simulated curves lie. We note, that it is crucial to consider median curves to study the underestimation of transmitted intensity values, since they are a measure of the most likely outcome of a simulation for a fixed number of simulated photon packages. Comparing the non-probabilistic solution in Fig. \ref{fig:comparison_Itrans_tau30} with the MC based intensity curves, we find that the simulated optical depth may result in severely underestimated values for the intensity, particularly if the stretching method is not used. In this case, even the usage of the method of enforced minimum scattering orders does not yield results that are close to the reference solution without simulating a higher number of photon packages. On the contrary, our results suggest, that using the stretching method leads to significantly better estimates and that the additional utilization of our proposed method (orange dashed curve) results in a further improvement of their quality. Altogether, enforcing a minimum scattering order seems to benefit the simulations twofold: First, the flux estimates are higher across all penetration angles. Second, the noise of the intensity estimates, which is inherent to the MC based approach, is smaller, which results in the overall narrower hatched orange region as compared to the blue region in the plot. 

   \begin{figure}
   \centering
   \includegraphics[width=\hsize]{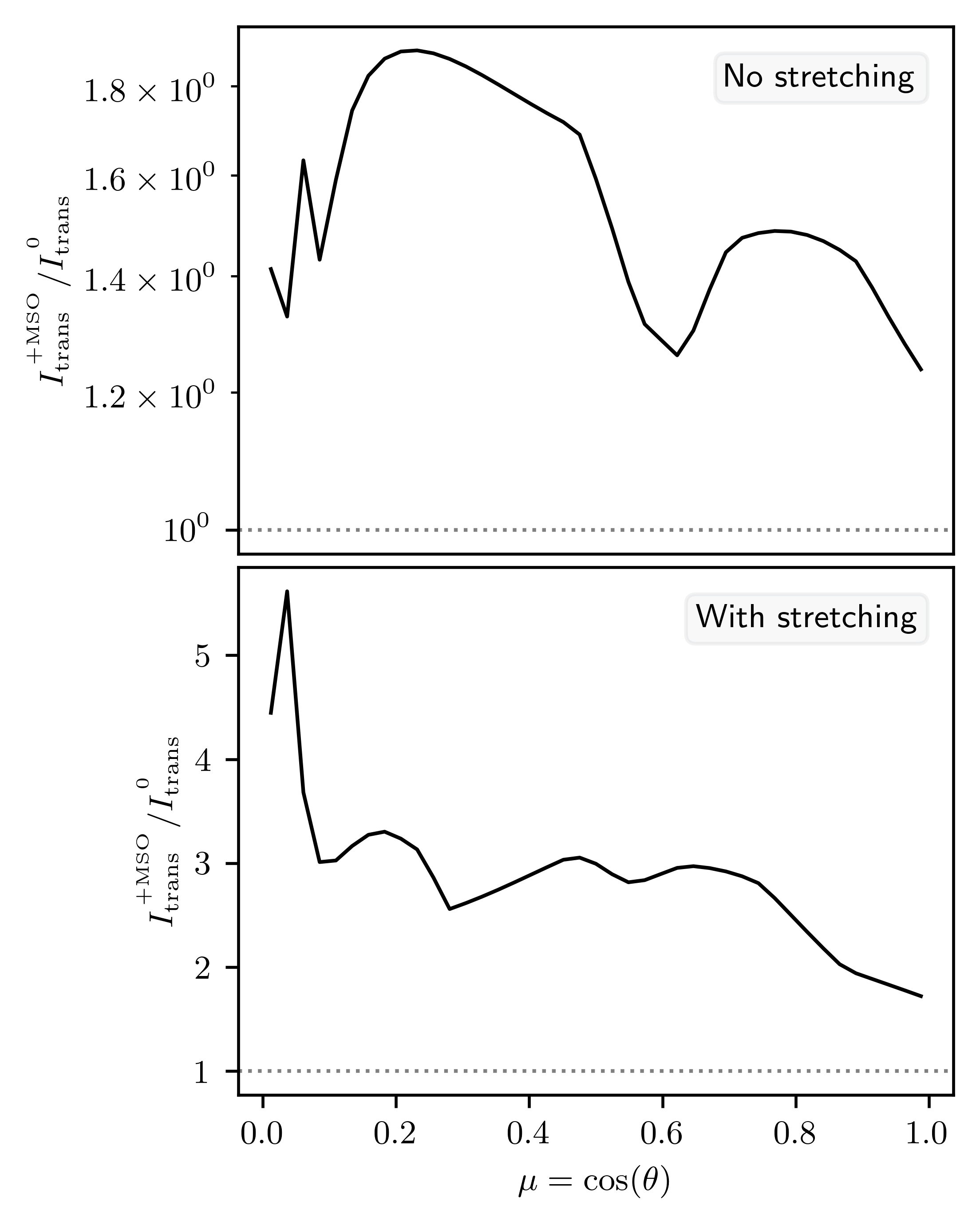}
      \caption{Ratios of transmission intensity curves $I^{\rm \,^{+MSO}}_{\rm trans} / I^{\rm \,^0}_{\rm trans}$ as function of the direction of penetration $\mu$, where $I^{\rm \,^{+MSO}}_{\rm trans}$ ($I^{\rm \,^0}_{\rm trans}$) denotes curve corresponding to simulations applying (not applying) the enforcement of a minimum scattering order. Values above one (gray dotted line) indicate a gain of intensity associated with the method. The upper (lower) plot shows the results for simulations that do not (do) additionally make use of a stretching method.}
         \label{fig:ratio_Itrans_tau30}
   \end{figure}
   
Figure \ref{fig:ratio_Itrans_tau30} shows the ratio of the median intensity curves $I^{\rm \,^{+MSO}}_{\rm trans} / I^{\rm \,^0}_{\rm trans}$ of simulations that utilize our proposed method ($I^{\rm \,^{+MSO}}_{\rm trans}$) and those that do not make use of it ($I^{\rm \,^0}_{\rm trans}$). In this plot, values above one (gray dotted line) indicate a gain of intensity associated with the usage of our method. The upper plot in Fig. \ref{fig:ratio_Itrans_tau30} shows the case when the stretching method is not used, and the lower plot that when it is used. Generally, we find that the intensity ratio is typically greater than one, indicating a gain of intensity associated with the usage of our proposed method. This is especially the case, when the stretching method is used, occasionally even leading to intensity ratios above $5$. When stretching is not used, the gain still amounts to values between $\sim\!20\!-\!90\,\%$. In further tests, where we reduced the number of simulated photon packages per simulation to $10^5$ or even to $10^4$, we found, that the severity of the underestimation of flux estimates clearly increases with lower photon package numbers. Likewise, an increase of the number of simulated photon packages has led to further improvement. Overall, the enforcement of the minimum scattering order therefore appears to be most effective, when the number of simulated photon packages generally does not suffice to ensure a proper representation of relevant photon package paths. In turn, this implies that its utilization can therefore greatly reduce the required number of simulated photon packages. 

\begin{figure}
   \centering
   \includegraphics[width=\hsize]{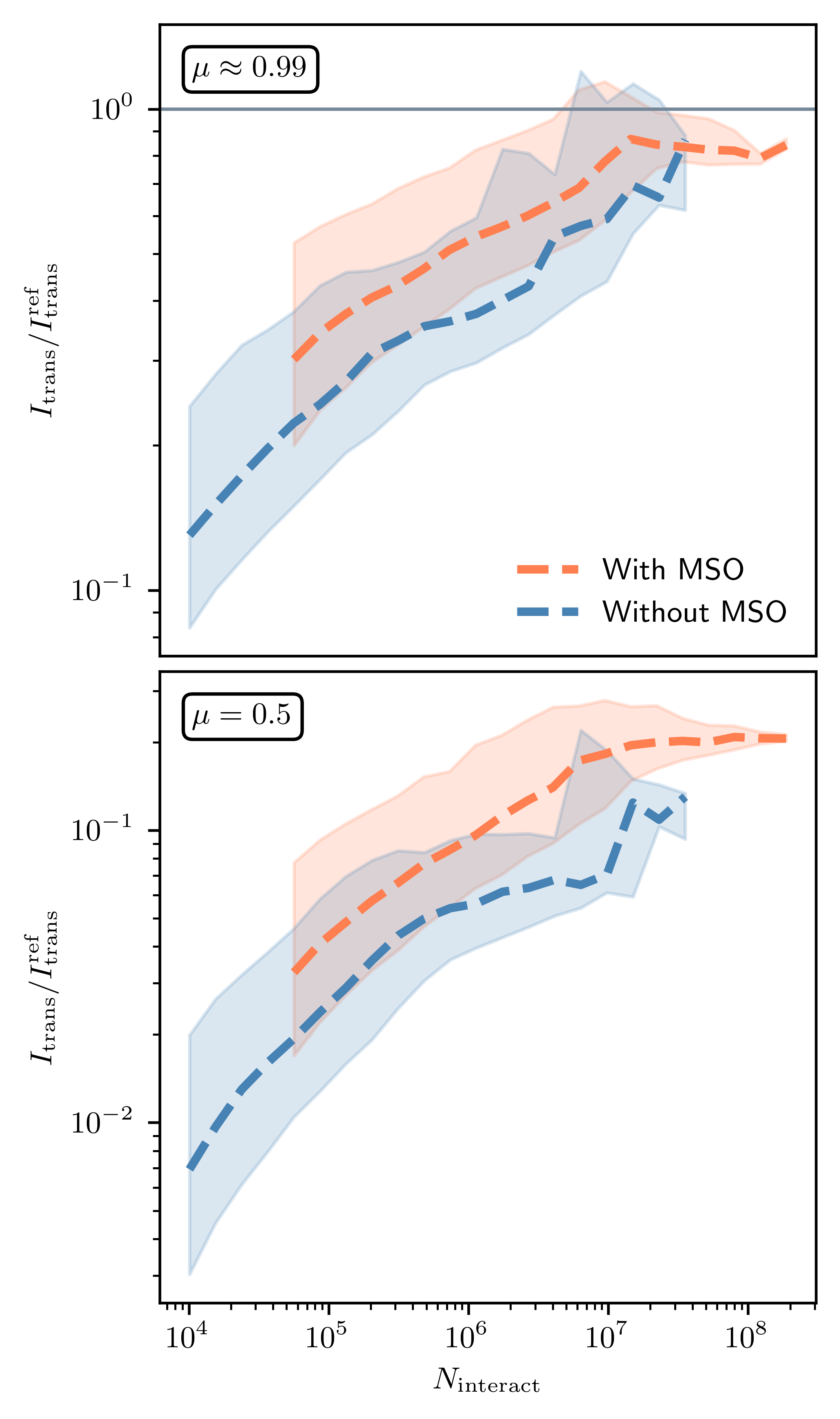}
      \caption{The ratio $I_{\rm trans}/I_{\rm trans}^{\rm ref}$ as a function of the number of simulated interaction events $N_{\rm interact}$, where $I_{\rm trans}^{\rm ref}$ denotes the value of the reference solution. The lines correspond to median curves, and the (color-coded) shaded areas encompass the central 50\,\% of corresponding data points. The upper (lower) plot shows the results for a penetration direction of $\mu \approx 0.99$ ($\mu=0.5$).}
         \label{fig:convergence_2dir}
   \end{figure}

The previous analysis has demonstrated that the application of our proposed method can lead to improved flux estimates, when keeping the number of simulated photon packages constant. However, the additional simulation of interaction events is also accompanied by an increase of computation time, which roughly scales with the number of simulated events. To assess the effect of its utilization on the computational demand, Fig. \ref{fig:convergence_2dir} shows the dependence of calculated intensity estimates as a function of the number of simulated events of interaction $N_{\rm interact}$. In particular, the upper plot shows median curves of the ratio $I_{\rm trans}/I_{\rm trans}^{\rm ref}$ for the direction $\mu\approx 0.99$, which roughly corresponds to the direction perpendicular to the slab surface, while assuming the usage of the stretching method. Here, $I_{\rm trans}^{\rm ref}$ is the respective linearly interpolated non-probabilistic reference solution. Additionally, the color-coded shaded areas in the plot mark the regions where the central 50\,\% (also called the interquartile range) of corresponding data points lie. We note that to generate these plots, the results of the previously mentioned $10^4$ performed simulations have been gathered in batches, assuming various sizes between 1 and $10^4$, and subsequently averaged. Additionally, for each batch, the corresponding numbers of simulated interactions of all simulations have been added up. The averaged intensity curves obtained in this way, with the corresponding numbers of interactions, were binned using 20 logarithmically sampled bins with regard to $N_{\rm interact}$, allowing for the determination of the interquartile ranges, as shown in this plot.

The upper plot exhibits a clear increase in the median ratio $I_{\rm trans}/I_{\rm trans}^{\rm ref}$ for both types of simulations. Simulations that utilize our method, though, achieve higher flux estimates for the same number of simulated events and are therefore closer to the reference solution. We note that data points in the area of the plot where $N_{\rm interact} \gtrsim 10^7$ originate from the averaging of large batches of simulations, of which there are only few. As a result, this procedure can be expected to lead to comparably poor statistics at very high values of $N_{\rm interact}$. This also explains the decrease in slope at $N_{\rm interact} \approx 10^7$ of both curves, which is an artifact that is expected for this analysis procedure. Apart from the clear improvement of the median curve, also the interquartile range of intensity values is much closer to the reference solution (see orange shaded area). Its width in the logarithmic plot, however, appears to be similar for both simulation types. 
The lower plot shows similar results, but now assuming a direction of $\mu=0.5$, which corresponds to a penetration angle of $60^\circ$. Both plots show similar trends, but the intensity is significantly more underestimated for the direction $\mu=0.5$ for all considered values of $N_{\rm interact}$. Nonetheless, the underestimation becomes less severe as $N_{\rm interact}$ increases, as expected, which is similar to the case of $\mu\approx 0.99$. This again suggests that for increasing penetration angles, the number of interactions events that needs to be simulated to achieve a certain level of quality increases. While the actual solution will be reached with or without the utilization of our method, given a sufficient number of photon packages are simulated, its usage appears to significantly reduce the required number of simulated interactions. 

Moreover, $N_{\rm interact}$ scales linearly with the computational time and is independent of the specific computer hardware, which makes it a suitable quantity for measuring the computational demand of MCRT simulations. Based on that, the plot also suggests that the usage of our method decreases the computational demand needed to achieve a certain level of quality of flux estimates. In particular, for a constant ratio of $I_{\rm trans}^{\rm ref}$, we find that fewer interactions (by about two to three times) have to be simulated for achieving the same quality, which corresponds to a likewise reduction of computation time, when enforcing our proposed minimum number of scattering orders. 

\subsection{Application: Embedded protoplanet in an edge-on disk}
\label{sec:application_disk}
In order to assess the benefits of our proposed method in an astrophysical setting, we implemented the method in the 3D Monte Carlo radiative transfer Code Mol3D \citep{Ober_2015}, a successor of MC3D \citep[][]{1999A&A...349..839W,Wolf_2003}. Mol3D uses a locally divergence free continuous absorption scheme of photon packages \citep{1999A&A...344..282L}, applies a method of immediate reemission according to a temperature corrected emission spectrum \citep{2001ApJ...554..615B}, and is particularly optimized for systems of extremely high optical depth by utilizing a large database of precalculated photon package paths \citep{2020A&A...635A.148K}. Furthermore, Mol3D uses the composite-biasing method \citep{2018ApJ...861...80C}, with which it has been equipped particularly for the purpose of this analysis. Details regarding Mol3D can be found in \citet{Ober_2015}.

The simulated system is composed of an edge-on circumstellar disk at a distance of $d=140\,$pc that harbors an embedded protoplanet. The disk density distribution is described by
\begin{equation}
\rho(r,z) = \frac{\Sigma (r)}{\sqrt{2\pi} h(r)}\cdot \exp\left[-\frac12 \left( \frac{z}{h(r)} \right)^2 \right], \label{eq:shakura_sunyaev}
\end{equation}
where $r$ is the radial distance to the center of the disk perpendicular to the $z$-axis and $z$ describes the distance to the mid-plane of the disk. $\Sigma (r)$ is the surface density and $h(r)$ the density scale height, which are given by \citep{1974MNRAS.168..603L,1998ApJ...495..385H}
\begin{equation}
\Sigma (r) = \Sigma_0 \left(\frac{r}{r_0}\right)^{\beta-\alpha}\cdot\exp\left[- \left( \frac{r}{r_0} \right)^{2+\beta-\alpha} \right]
\end{equation}
and
\begin{equation}
h(r) = h_0\left( \frac{r}{r_0} \right)^\beta,
\end{equation}
where variables with subscript 0 denote reference values evaluated at the reference radius $r_0$, $\alpha$ is the compactness parameter, and $\beta$ the flaring parameter of the disk. The total disk mass $M_{\rm disk}=10^{-2}\,{\rm M}_\odot$ includes the mass of dust and gas. Generally, the chosen values of all model parameters of the circumstellar disk correspond to commonly observed values that are typical for disks around T Tauri stars \citep{2009ApJ...700.1502A}. The dust is composed of spherical dust grains which themselves consist of $62.5\,\%$ silicate and $37.5\,\%$ graphite, the latter with a ratio of $\text{parallel}{:}\text{ortho}=1{:}2$ \citep{1993ApJ...414..632D}, for which we use mixed optical properties \citep{1984ApJ...285...89D,1993ApJ...402..441L,2001ApJ...548..296W}. The grains have a bulk density of $\rho_\text{bulk}=2.5\,\text{g}\,\text{cm}^{-3}$ and radii $a$ between 5 and $250\,$nm, that follow the distribution ${\rm d}n \sim a^{-3.5} {\rm d}a$ \citep{1977ApJ...217..425M}. Mean optical properties of these dust grains were calculated \citep{Wolf_2003} and used throughout our simulations. In these simulations, we assume Mie theory for the description of simulated scattering events. 

The embedded protoplanet orbits the center of the edge-on circumstellar disk at a distance of $r_{\rm p}=20\,$au and is placed between the observer and the disk center inside the mid-plane. It has a surface temperature of 2000\,K and a luminosity of $L_{\rm p}=10^{-3}\,{\rm L}_\odot$, which is consistent with evolutionary tracks of planets \citet{1997ApJ...491..856B}. A list of all relevant model parameters can be found in Tab. \ref{tbl:parameters}.

\begin{table}
\begin{center}
\begin{tabular}{lcc}
\toprule
Description & Parameter & Value\\
\cmidrule{1-3}
\multicolumn{3}{c}{\underline{Star}}\\
Distance                   & $d$ / pc                                &      140         \vspace{4px}\\
\multicolumn{3}{c}{\underline{Disk}}\\
Total mass                  & $M_\text{disk}$ / $\text{M}_\odot$      & $10^{-2}$        \\
Inner radius          & $r_\text{in}$ /  $\text{au}$            & $0.07$           \\
Outer radius          & $r_\text{out}$ /  $\text{au}$           & $300$            \\ 
Reference radius           & $r_\text{0}$ /  $\text{au}$             & $100$            \\
Reference scale height     & $h_\text{0}$ /  $\text{au}$             & $7$              \\
Compactness parameter & $\alpha$                                &       1.8        \\ 
Flaring parameter     & $\beta$                                 &       1.1        \\ 
Inclination                & $i$ / deg                               &       90         \\ 
Dust to gas ratio          &                                         &  1\,$:$\,100     \vspace{4px}\\ 
\multicolumn{3}{c}{\underline{Planet}}\\
Luminosity      & $L_{\rm p}$ /  $\text{L}_\odot$                            &       $10^{-3}$       \\  
Surface temperature      & $T_{\rm p}$ / K                            &       2000       \\  
Orbital radius            & $r_{\rm p}$ / au                           &       20         \\  
\bottomrule
\end{tabular}
\caption{Model parameters of the protoplanetary disk and the embedded planet. For details, see Sect. \ref{sec:application_disk}.}
\label{tbl:parameters}
\end{center}
\end{table}

Here, we consider two types of simulations for which we determine the observed total (specific) flux density that originates from the embedded planet. One type uses our proposed method of enforced minimum scattering order, while the other is not. Since the planet is embedded deeply inside the disk, we can expect the optical depth to be high in all directions, while the albedo value is rather small for the considered wavelength range. Therefore, we use the expression for the minimum scattering order $m$ in Eq. \eqref{eq:m_999} for the albedo-dominated case for our simulations. We note, that due to the general lack of a non-probabilistic reference solution for such a system, it is not possible to access the performance of MCRT simulations in the same manner as has been done for the case of the infinite plane-parallel slab (see Fig. \ref{fig:comparison_Itrans_tau30}). Therefore, we only focus on the flux gain associated with the usage of our proposed method. 

\begin{figure}
   \centering
   \includegraphics[width=\hsize]{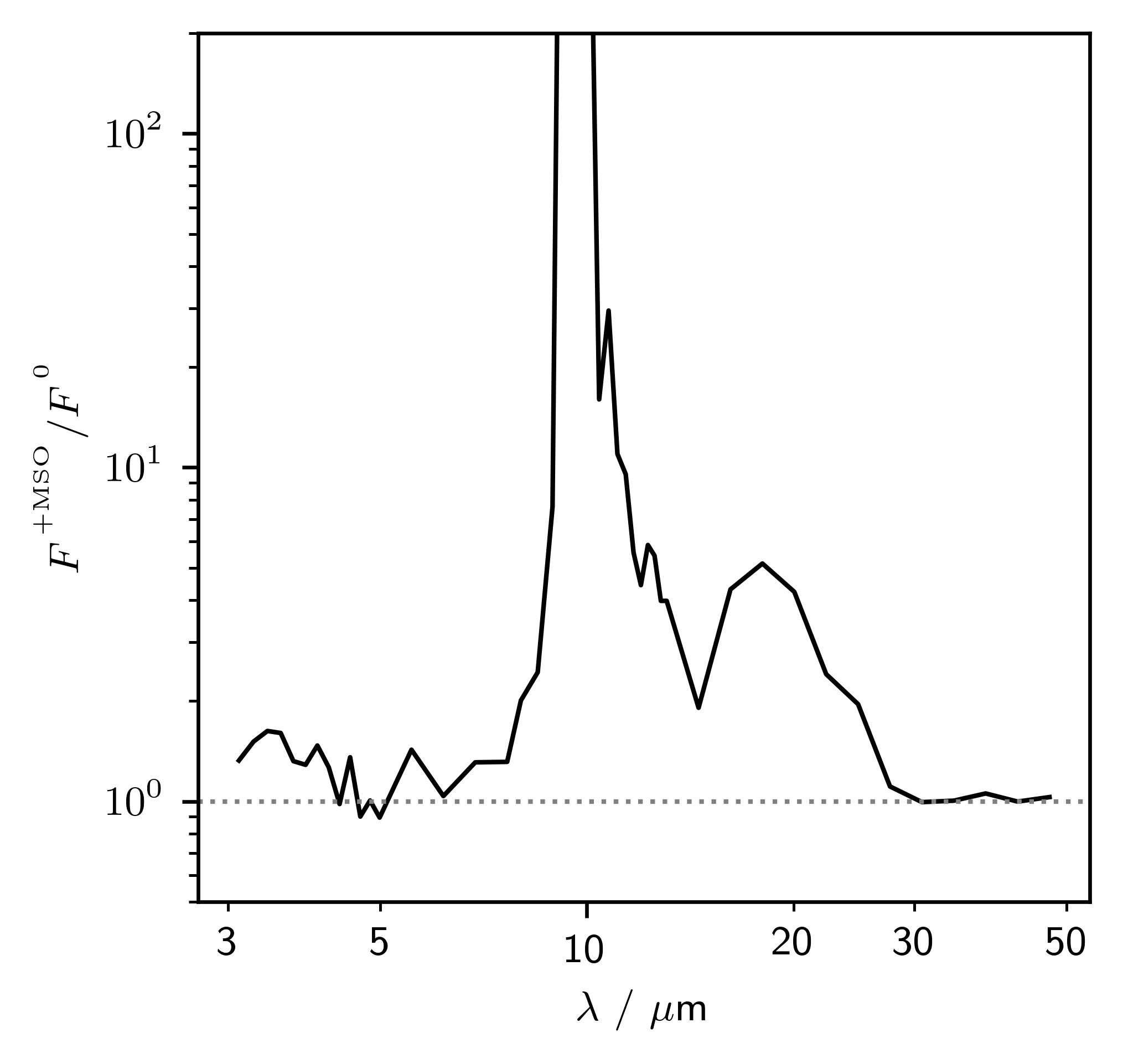}
      \caption{Results from simulations of an embedded protoplanet in an edge-on circumstellar disk, showing the ratio of two median flux density curves ($F^{\rm \,^{+MSO}} / F^{\rm \,^0}$) corresponding to radiation originating from the planet. Here, $F^{\rm \,^{+MSO}}$ ($F^{\rm \,^0}$) denotes the flux density curve obtained from simulations using (not using) the method of enforced minimum scattering order. Values above one (gray dotted line) indicate a gain in flux density associated with the usage of the method.
      }
         \label{fig:mol3d_results}
   \end{figure}

In total, we performed 400 simulations per type, each for 45 wavelengths $\lambda$ in the mid-infrared range between $\sim$3 and 50\,$\mu$m using $10^5$ photon packages per simulation and wavelength. Subsequently, we determined the ratio of the two resulting median flux density curves $F^{\rm \,^{+MSO}} / F^{\rm \,^0}$, where $F^{\rm \,^{+MSO}}$ ($F^{\rm \,^0}$) denotes the curve obtained from simulations using (not using) the method of enforced minimum scattering order. The results of these simulations are shown in Fig. \ref{fig:mol3d_results}. The plot shows, that for the majority of considered wavelengths, the usage of our proposed method leads to an increase of the determined flux estimate by a factor that partly even exceeds the value of $100$. Similarly to the findings in Sect. \ref{sec:application_slab}, we additionally find in this setup that the flux gain associated with the usage of our proposed method seems to increase as the expected level of flux decreases. Further tests were conducted in which we varied the surface temperature and luminosity of the planet as well as its orbital position, both radially and azimuthally, which only provided further evidence in support of this conclusion.

\section{Discussion and implications}
\label{sec:discussions}

In this section, we give the reasons for using our proposed method as a simple way to implement the means of providing a minimum quality during the step of flux determination in a MCRT simulation. We additionally address potential issues concerning the applicability of the method and the interpretation of the results in Eqs. \eqref{eq:m_general} and \eqref{eq:m_999}. Altogether, we argue in favor of using a minimum scattering order $m$ that is at least as high as the expression derived in Eq. \eqref{eq:m_999} for MCRT simulations of all systems which, for instance, contain regions of high optical depths, or systems that contain a radiation source, which is embedded inside an interacting medium. 

 While the actual SOD of a simulated system depends on its particular properties, our method of the enforced minimum scattering order is aimed at avoiding the consideration of the full complexity and gives an estimate for it in terms of a minimum scattering order $m$, which can readily be used throughout a simulation. To clarify the role of $m$ in such a simulation, we next discuss three changes to the setup of the 1D slab, which would lead to a shift and reshaping of the SOD; while these changes are specific, the drawn conclusions are generally valid. First, we discuss the effects of placing the radiation source inside the slab, which shifts the SOD toward lower scattering orders. Second, we consider the difference between the approximate solution for the SOD (Eq. \ref{eq:In_using_exp_approximation}) and the actual solution, which is shifted toward higher scattering orders. Third, we consider the transition from the 1D to the 3D case, which leads to a shift and reshaping of the SOD. We discuss these changes to the SOD while assuming the case of $A<1$, as this applies to the majority of materials in astrophysics, which, in accordance with Eq. \eqref{eq:m_999}, correspond to a minimum scattering order in the albedo limited case of $m=-7/\ln{A} - 1$. The reasoning, however, can be applied to the optical depth-dominated case as well.

By placing the radiation source inside the slab rather than outside, on the side that is not facing the detector, the optical depth between the detector and the source of radiation decreases. One crucial consequence of the reduced optical depth is that photon packages of lower scattering orders gain in relevance regarding their contribution to the total detected intensity, tendentially leading to a shift of the SOD toward lower scattering orders. This could potentially result in a situation whereby $m$ is overestimated, such that the coverage of scattering orders would exceed $p>99.9\%$. When making use of our proposed method, this by itself is not an unfavorable result, as it would simply increase the quality of the determined flux estimates. If, however, the required simulation time for this is expensive, the achieved gain in accuracy may not be worth the accompanied increase in simulation time. Therefore, a shift of the SOD toward lower scattering orders does not result in a lowering of the quality of estimated flux values, when using our proposed method of enforced minimum scattering order.  

If, on the other hand, the SOD is shifted toward higher scattering orders, as can be expected (see Sect. \ref{sec:SOD}) for the actual distribution, compared to its approximation in Eq. \eqref{eq:In_using_exp_approximation}, the proposed value of $m$ may be below what would be required to achieve the desired coverage of $p=99.9\%$. As a result, the derived flux values may still be underestimated, however, due to the enforced contribution of scattering orders up to the order of $m$, the level of underestimation can be expected to be lower. Therefore, the usage of the method would at the very least provide an improvement of quality of determined flux values (as has been shown in Sections \ref{sec:application_slab} and \ref{sec:application_disk}) and offset the scattering orders of photon packages closer to the relevant scattering order interval. 

The same reasoning holds when transitioning from the 1D to the 3D scenario. This can be understood, for instance, by assuming a slightly changed setup and comparing the scenario of a central radiation source that is located inside a 1D slab of half transverse width $b$ with that of a 3D sphere of radius $b$, both in units of optical depth. Since in the 1D case, the photon package path is confined, such that it always moves toward or away from the center, its movement in the radial direction leads to an average shift of its position of $\Delta \tau_r=1$ between two consecutive events of interaction. In the 3D case and in the regime of high optical depths, the average displacement of the radial position of the photon package is comparably smaller. In particular, if the photon package is already at a radial optical depth position of $\tau_r\gg 1$, the average displacement reduces to $\Delta \tau_r=0.5$. The reduction of average radial displacement is a result of the fact, that the direction of the photon package in the 1D case always is parallel to the radial direction, which it is not in general in the 3D case. As a result, the SOD in the 3D case can be expected to be, firstly, narrower compared to that of the 1D case, and secondly, shifted toward higher scattering orders, since the radial displacement of photon packages is delayed due to the smaller radial pacing at high optical depths. Therefore, we can conclude, that despite the SOD exhibiting a different shape and being shifted, using the minimum scattering order $m$ from Eq. \eqref{eq:m_999} for 3D MCRT simulations can still be expected to elevate the quality of determined flux values (compare with Sections \ref{sec:application_slab} and \ref{sec:application_disk}) and serve as a lower boundary for the measure of the complexity of the underlying SOD.

Our findings clearly suggest that using the quantity $m$ from Eqs. \eqref{eq:m_general} and \eqref{eq:m_999} as a measure of the minimum scattering order for the proposed method of enforced minimum scattering order generally leads to more reliable and higher flux values in both the 1D and the 3D cases. This is particularly relevant in the regime of high optical depths, which can be explained as a result of the alleviation of the scattering order problem. This has implications not only for simulations in which photon packages themselves are used to construct observables, but also when relying on secondary results derived from their simulation. One such example would be the utilization of a raytracing scheme for deriving flux maps, which is based on a scattering source function that is derived during a MCRT simulation. For such a case, our findings clearly suggest that this source function would also be lacking in terms of the contribution of photon packages, which would ultimately manifest itself in underestimated flux values; the latter could likewise be improved by the usage of our proposed method. Moreover, we note that while it is possible to adjust the method for the estimation of $m$ by assuming a different coverage $p$ of the SOD, we generally recommend using high values close to unity. Ultimately, it is still necessary to limit its value.

Furthermore, our results have implications for the performance of other optimization methods used to improve MCRT in the regime of high optical depths. This is the case, for instance, in the context of biasing methods that affect the determination of photon package paths, such as stretching methods \citep[e.g.,][]{2018ApJ...861...80C}, which increase the average distance between two consecutive points of interaction of a simulated photon package. As we discussed previously \citep{2021A&A...645A.143K}, the usage of such a stretching method, despite its great potential benefits, has its limitations, which can be the result of a too strong decrease of sampled scattering orders of simulated photon packages. This can lead to a scenario in which the SOD is not sufficiently well represented as the relevant scattering order intervals are practically not reached due to the usage of the stretching method. This may very well explain the improvement of flux estimates that was associated with the usage of our method in the embedded protoplanet case described in Sect. \ref{sec:application_disk}. In further tests, where we performed the same simulations but did not apply the stretching method, we found no difference between simulations that enforced a minimum number of scattering orders and those that did not. This can be explained by the fact that simulated photon packages that are emitted by the planet cannot escape without a high number of interactions. 

The findings of this study therefore suggest that the usage of a stretching method, which results in a sampling of scattering orders well below the minimum scattering order, $m$, can be considered as an indication for too high stretching factors. Moreover, the best suited stretching methods should take into account the albedo of the medium, as it strongly shapes the complexity of the simulation, see Sect. \ref{sec:SOD}. However, our results do suggest that in the albedo-dominated case, namely, low albedo and high optical depth, the usage of a method that increases the overall pace with which a simulated photon package is transported, generally offers a great potential for MCRT simulations, if used correctly, as can be seen in Fig. \ref{fig:comparison_Itrans_tau30}. This implies that the best-suited method to overcome the problem of high optical depths and the flux underestimation problem seems to utilize a stretching technique combined with a method to ensure a sufficient number of simulated interactions events. 

As a precautionary note, it is worth acknowledging that when dealing with more complex systems beyond the scope of our current consideration, it may be generally difficult to decide whether the problem is albedo-dominated or optical depth-dominated. This could for instance be the case, if the albedo is direction-dependent or the optical properties of the simulated system vary across the medium. However, as long as the optical depth is generally high, we believe that it is worth considering the usage of a stretching technique combined with the enforcement of at least as many scattering orders $m$ as derived for the albedo-dominated case, see Eq. \eqref{eq:m_999}. Furthermore, if this is leading to higher flux estimates, it would be crucial to assess whether a further increase of $m$ would yield even higher estimates. 

Moreover, we note that in the case of a low albedo values and depending on the particular MCRT implementation, the typical number of simulated scattering events may also exceed the minimum scattering order for a large portion of simulated scattering orders, while estimated flux values are still underestimated. This may hint at another problem, which lies in a suboptimal method for path determination, which in this case leads to an oversampling of paths of high scattering orders and an insufficient representation of paths of low scattering orders. Thus, it may be beneficial to make use of a different MCRT implementation and apply, for instance, better suited biasing techniques. Additionally, the question regarding a concept of a maximum scattering order may also arise, such that photon packages that have undergone a certain number of scattering events may safely be removed from the model space. The knowledge of such a maximum scattering order may then be used to increase the efficiency of MCRT simulations by spending less time on the simulation of insignificantly and more on the simulation of significantly contributing photon packages. Overall, we find that it would be highly desirable to design an adaptive stretching method in the future, which takes into account the findings of this study in order to adjust the stretching as needed to ensure the most reliable flux estimates.

\section{Summary}
\label{sec:summary}
In this study, we analyzed the scattering order distribution (SOD) of transmitted photon packages, which traversed a 1D slab that is illuminated from one side. On the basis of this testbed, we derived an approximate solution for the SOD and studied its implications for the complexity of Monte Carlo radiative transfer (MCRT) simulations, as described in Sects. \ref{sec:SOD} and \ref{sec:complexity}. Here, we find further evidence in favor of the conjecture that the scattering order problem constitutes one of the issues that lies at the core of MCRT simulations, particularly in the regime of high optical depths. Moreover, we identified two different cases that can dominate the complexity of a MCRT simulation at high optical depths with notably distinct properties, i.e., the albedo-dominated case ($|\ln{\lambda_1}|\ll |\ln{A}|$) and the optical depth-dominated case ($|\ln{A}|\ll |\ln{\lambda_1}|$). On this basis, we derived the following three properties of the approximate SOD, all of which are in agreement with previously made predictions \citep[see][]{2021A&A...645A.143K}. First, the peak of the SOD at the scattering order $n_{\rm max}$ can be used as a measure of the average scattering order of the distribution. Second, the value of $n_{\rm max}$ scales approximately quadratically with the optical depth $\tau_{\rm max}$ in the optical depth-dominated case, while being limited by $n_{\rm max}\lesssim -1/\ln{A}$ independent of the value of the optical depth in the albedo-dominated case. Third, the apparent width of the SOD in the double logarithmic plot converges in the limit of $\tau_{\rm max}\rightarrow \infty$, which stresses the severity of the scattering order problem, especially in the optical depth-dominated case. 

Furthermore, we used the SOD approximation to derive an estimate for the minimum scattering order $m$, which is required to achieve a certain coverage of the SOD and the total transmitted intensity (see Sect. \ref{sec:minimum_scattering_order}). We thus propose a method of enforced minimum scattering order that ensures a minimum scattering order of simulated photon packages of at least the derived value of $m$. Our proposed method aims at increasing the reliability of flux estimates, by alleviating the scattering order problem, which thus serves as a simple to implement means of providing a minimum quality both in 1D and 3D simulations. 

We then tested the method by applying it in two different simulated systems: 1) an infinite plane-parallel slab (see Sect. \ref{sec:application_slab}) and 2) a protoplanet embedded in an edge-on circumstellar disk (see Sect. \ref{sec:application_disk}). In both systems, the enforcement of the minimum scattering order led to an alleviation of the underestimation of flux and its effectiveness seemed to increase for lower observed values of flux. Overall, we find that this concept bares great potential for an improved flux determination, particularly in the regime of high optical depths and in combination with a photon path length stretching technique, as described in Sect. \ref{sec:discussions}. Finally, we have discussed the possibility of more advanced optimization techniques for the flux determination step that build on our findings regarding the SOD and minimum scattering order. With the help of such techniques, it might then be possible to overcome the core problems in the regime of high optical depths and as a result maximize the overall potential of MCRT simulations.

\section*{ORCID iDs}
A. Krieger \orcidlink{0000-0002-3639-2435}
\href{https://orcid.org/0000-0002-3639-2435}
     {https://orcid.org/0000-0002-3639-2435}\\
S. Wolf \orcidlink{0000-0001-7841-3452}
\href{https://orcid.org/0000-0001-7841-3452}
     {https://orcid.org/0000-0001-7841-3452}

\begin{acknowledgements}
         We thank all the members of the Astrophysics Department Kiel for helpful discussions and remarks. 
         We acknowledge the support of the DFG priority program SPP 1992 "Exploring the Diversity of Extrasolar Planets (WO 857/17-2)".
\end{acknowledgements}

\bibliography{literature}

\begin{thebibliography}{27}
\expandafter\ifx\csname natexlab\endcsname\relax\def\natexlab#1{#1}\fi

\bibitem[{{Andrews} {et~al.}(2009){Andrews}, {Wilner}, {Hughes}, {Qi}, \&
  {Dullemond}}]{2009ApJ...700.1502A}
{Andrews}, S.~M., {Wilner}, D.~J., {Hughes}, A.~M., {Qi}, C., \& {Dullemond},
  C.~P. 2009, \apj, 700, 1502

\bibitem[{{Baes} {et~al.}(2016){Baes}, {Gordon}, {Lunttila}, {Bianchi},
  {Camps}, {Juvela}, \& {Kuiper}}]{2016A&A...590A..55B}
{Baes}, M., {Gordon}, K.~D., {Lunttila}, T., {et~al.} 2016, \aap, 590, A55

\bibitem[{{Bjorkman} \& {Wood}(2001)}]{2001ApJ...554..615B}
{Bjorkman}, J.~E. \& {Wood}, K. 2001, \apj, 554, 615

\bibitem[{{Burrows} {et~al.}(1997){Burrows}, {Marley}, {Hubbard}, {Lunine},
  {Guillot}, {Saumon}, {Freedman}, {Sudarsky}, \&
  {Sharp}}]{1997ApJ...491..856B}
{Burrows}, A., {Marley}, M., {Hubbard}, W.~B., {et~al.} 1997, \apj, 491, 856

\bibitem[{{Camps} \& {Baes}(2018)}]{2018ApJ...861...80C}
{Camps}, P. \& {Baes}, M. 2018, \apj, 861, 80

\bibitem[{{Camps} \& {Baes}(2020)}]{2020A&C....3100381C}
{Camps}, P. \& {Baes}, M. 2020, Astronomy and Computing, 31, 100381

\bibitem[{{di Bartolomeo} {et~al.}(1995){di Bartolomeo}, {Barbaro}, \&
  {Perinotto}}]{1995MNRAS.277.1279D}
{di Bartolomeo}, A., {Barbaro}, G., \& {Perinotto}, M. 1995, \mnras, 277, 1279

\bibitem[{{Draine} \& {Lee}(1984)}]{1984ApJ...285...89D}
{Draine}, B.~T. \& {Lee}, H.~M. 1984, \apj, 285, 89

\bibitem[{{Draine} \& {Malhotra}(1993)}]{1993ApJ...414..632D}
{Draine}, B.~T. \& {Malhotra}, S. 1993, \apj, 414, 632

\bibitem[{{Dullemond} {et~al.}(2012){Dullemond}, {Juhasz}, {Pohl}, {Sereshti},
  {Shetty}, {Peters}, {Commercon}, \& {Flock}}]{2012ascl.soft02015D}
{Dullemond}, C.~P., {Juhasz}, A., {Pohl}, A., {et~al.} 2012, {RADMC-3D: A
  multi-purpose radiative transfer tool}

\bibitem[{{Gordon} {et~al.}(2017){Gordon}, {Baes}, {Bianchi}, {Camps},
  {Juvela}, {Kuiper}, {Lunttila}, {Misselt}, {Natale}, {Robitaille}, \&
  {Steinacker}}]{2017A&A...603A.114G}
{Gordon}, K.~D., {Baes}, M., {Bianchi}, S., {et~al.} 2017, \aap, 603, A114

\bibitem[{{Hartmann} {et~al.}(1998){Hartmann}, {Calvet}, {Gullbring}, \&
  {D'Alessio}}]{1998ApJ...495..385H}
{Hartmann}, L., {Calvet}, N., {Gullbring}, E., \& {D'Alessio}, P. 1998, APJ,
  495, 385

\bibitem[{{Krieger} \& {Wolf}(2020)}]{2020A&A...635A.148K}
{Krieger}, A. \& {Wolf}, S. 2020, A\&A, 635, A148

\bibitem[{{Krieger} \& {Wolf}(2021)}]{2021A&A...645A.143K}
{Krieger}, A. \& {Wolf}, S. 2021, \aap, 645, A143

\bibitem[{{Laor} \& {Draine}(1993)}]{1993ApJ...402..441L}
{Laor}, A. \& {Draine}, B.~T. 1993, \apj, 402, 441

\bibitem[{{Lucy}(1999)}]{1999A&A...344..282L}
{Lucy}, L.~B. 1999, \aap, 344, 282

\bibitem[{{Lynden-Bell} \& {Pringle}(1974)}]{1974MNRAS.168..603L}
{Lynden-Bell}, D. \& {Pringle}, J.~E. 1974, MNRAS, 168, 603

\bibitem[{{Mathis} {et~al.}(1977){Mathis}, {Rumpl}, \&
  {Nordsieck}}]{1977ApJ...217..425M}
{Mathis}, J.~S., {Rumpl}, W., \& {Nordsieck}, K.~H. 1977, \apj, 217, 425

\bibitem[{{Min} {et~al.}(2009){Min}, {Dullemond}, {Dominik}, {de Koter}, \&
  {Hovenier}}]{2009A&A...497..155M}
{Min}, M., {Dullemond}, C.~P., {Dominik}, C., {de Koter}, A., \& {Hovenier},
  J.~W. 2009, \aap, 497, 155

\bibitem[{{Ober} {et~al.}(2015){Ober}, {Wolf}, {Uribe}, \& {Klahr}}]{Ober_2015}
{Ober}, F., {Wolf}, S., {Uribe}, A.~L., \& {Klahr}, H.~H. 2015, \aap, 579, A105

\bibitem[{{Reissl} {et~al.}(2016){Reissl}, {Wolf}, \&
  {Brauer}}]{2016A&A...593A..87R}
{Reissl}, S., {Wolf}, S., \& {Brauer}, R. 2016, \aap, 593, A87

\bibitem[{{Roberge}(1983)}]{1983ApJ...275..292R}
{Roberge}, W.~G. 1983, \apj, 275, 292

\bibitem[{{Robitaille}(2010)}]{2010A&A...520A..70R}
{Robitaille}, T.~P. 2010, \aap, 520, A70

\bibitem[{{Weingartner} \& {Draine}(2001)}]{2001ApJ...548..296W}
{Weingartner}, J.~C. \& {Draine}, B.~T. 2001, \apj, 548, 296

\bibitem[{Wolf(2003)}]{Wolf_2003}
Wolf, S. 2003, The Astrophysical Journal, 582, 859

\bibitem[{{Wolf} {et~al.}(1999){Wolf}, {Henning}, \&
  {Stecklum}}]{1999A&A...349..839W}
{Wolf}, S., {Henning}, T., \& {Stecklum}, B. 1999, \aap, 349, 839

\bibitem[{{Yusef-Zadeh} {et~al.}(1984){Yusef-Zadeh}, {Morris}, \&
  {White}}]{1984ApJ...278..186Y}
{Yusef-Zadeh}, F., {Morris}, M., \& {White}, R.~L. 1984, \apj, 278, 186

\end{thebibliography}
\bibliographystyle{aa}



\end{document}